\documentclass[cameraready]{IEEEtran}
\IEEEoverridecommandlockouts

\usepackage{cite}
\usepackage{amsmath,amssymb,amsfonts}
\usepackage{algorithmic}
\usepackage{graphicx}
\usepackage{textcomp}
\usepackage{xcolor}
\usepackage{booktabs}
\usepackage{url}
\usepackage{hyperref}

\hypersetup{
    pdfauthor={Houmin Sun, Zi Hu, Linxi Li, Yechen Wang, Liwei Jin, Carsten Maple, Ming Li},
    pdftitle={Investigating Codec-Internal Latent Audio Watermarking for Neural Codec Robustness},
    pdfsubject={Audio Watermarking},
    pdfkeywords={audio watermarking, audio codec, robustness, latent embedding, RVQ},
    pdfproducer={LaTeX},
    pdfcreator={pdfTeX}
}

\def\BibTeX{{\rm B\kern-.05em{\sc i\kern-.025em b}\kern-.08em
    T\kern-.1667em\lower.7ex\hbox{E}\kern-.125emX}}

\begin{document}

\title{Investigating Codec-Internal Latent Audio Watermarking for Neural Codec Robustness}

\author{
\IEEEauthorblockN{
Zi Hu\IEEEauthorrefmark{1}\textsuperscript{*},
Houmin Sun\IEEEauthorrefmark{1}\textsuperscript{*},
Linxi Li\IEEEauthorrefmark{3},
Yechen Wang\IEEEauthorrefmark{3},
Liwei Jin\IEEEauthorrefmark{3},
Carsten Maple\IEEEauthorrefmark{4}\textsuperscript{\textdagger},
Ming Li\IEEEauthorrefmark{2}\textsuperscript{\textdagger}
}

\IEEEauthorblockA{
\IEEEauthorrefmark{1}Digital Innovation Research Center,
Duke Kunshan University, Kunshan, China\\
\IEEEauthorrefmark{2}School of Artificial Intelligence,
The Chinese University of Hong Kong, Shenzhen, China\\
\IEEEauthorrefmark{3}OfSpectrum, Inc., Los Angeles, USA\\
\IEEEauthorrefmark{4}University of Warwick, Coventry, United Kingdom\\
Email: ming.li.cuhksz@gmail.com
}

\thanks{\textsuperscript{*}Equal contribution.}
\thanks{\textsuperscript{\textdagger}Corresponding author: Ming Li, Carsten Maple.}
}

\maketitle

\begin{abstract}
Neural audio codecs are challenging transformations for audio watermarking
because they re-encode, quantize, and resynthesize speech. This paper
investigates continuous latent-space watermarking for codec robustness. Instead
of adding a watermark only to the waveform or spectrogram, we embed a 32-bit
message into the continuous latent representation of a codec-like speech
autoencoder. The pipeline uses a SEANet-style encoder-decoder, a
Conformer-based message embedder, RVQ-guided latent decomposition, and a
latent-domain detector trained under signal-processing and neural-codec
transformations. Rather than proposing a final universal watermarking baseline,
we characterize the trade-offs that appear when the watermark carrier is moved
before neural decoding. On 48 kHz speech, EnCodec-aware training improves
EnCodec-24k bit accuracy from 78.8\% to 95.6\% and 97.1\%, while PESQ
decreases from 3.727 to 3.514 and 3.427.
\end{abstract}


\section{Introduction}
Recent progress in AI speech synthesis has made high-quality synthetic speech increasingly accessible~\cite{ren2020fastspeech2,wang2023valle}, raising urgent needs for reliable provenance tracking and misuse detection. Audio watermarking offers a practical mechanism for embedding hidden identity or message bits into generated speech, allowing later verification without requiring external metadata. To preserve perceptual quality, most recent deep-learning-based watermarking systems embed watermarks in the waveform or time-frequency domain with carefully constrained perturbations~\cite{chen2023wavmark,sanroman2024audioseal,singh2024silentcipher}. With differentiable or simulated augmentation, these encoder-detector systems can make watermarks imperceptible while remaining robust to common signal-processing attacks such as noise, filtering, resampling, reverberation, and compression.

However, neural audio codecs introduce a qualitatively different and more destructive transformation. Unlike simple filtering or lossy compression, modern codecs first map the waveform into a learned latent representation, quantize this representation, and then synthesize the output waveform through a neural decoder~\cite{zeghidour2021soundstream,defossez2022encodec,yang2023hificodec}. This analysis-quantization-synthesis process can discard weak watermark traces that are not aligned with the codec representation. In particular, RVQ maps continuous features to finite codebook entries, so small waveform- or spectrogram-level perturbations may be absorbed by quantization or reshaped by the decoder without preserving the message. As neural codecs become common in speech generation, distribution, and editing pipelines~\cite{wang2023valle,zhang2023speechtokenizer}, improving watermark robustness under codec transformations becomes a central challenge.

Motivated by this observation, we explore watermark embedding directly in the latent representation of a codec-like speech autoencoder. The key intuition is that a watermark inserted at a deeper representational level can be made more consistent with the factors that the decoder uses to reconstruct speech, rather than existing only as a fragile surface-level perturbation. In such a space, the watermark can interact with content features before waveform synthesis, potentially making it less likely to be removed by subsequent codec operations. 

We explore several ways of inserting message information into this latent space. Early attempts used multiplicative latent masking and a hard RVQ bottleneck in the reconstruction path, but both made optimization fragile: the former altered the scale of content-bearing latent features, while the latter forced speech reconstruction and message preservation through discontinuous codebook assignments. The reported system therefore uses a bounded additive residual in continuous latent space and keeps RVQ as a guidance and decomposition mechanism rather than as the mandatory reconstruction bottleneck. Rather than claiming a final optimized watermarking system, we use this pipeline to characterize the quality-robustness trade-off that appears when message recovery is trained through a continuous latent reconstruction path.

Our proposed pipeline encodes speech with a SEANet-like encoder, embeds a message-conditioned watermark in the continuous latent representation, reconstructs the waveform with a neural decoder, and trains a latent-domain detector to recover the message after codec and signal-processing transformations. We evaluate both perceptual quality and watermark robustness, focusing especially on the trade-off between imperceptibility and survival under neural codec attacks. The main contributions of this work are:
\begin{itemize}
\item We formulate codec-robust speech watermarking as an investigation of where and how message information can be embedded inside continuous latent representations of codec-like autoencoders.

\item We examine a set of continuous latent embedding mechanisms, including multiplicative masking, bounded residual injection, RVQ-guided decomposition, and protected front-layer embedding, as design attempts for understanding how watermark information can be inserted before neural decoding.

\item We empirically analyze the trade-off among codec robustness, perceptual quality, reconstruction backbone capacity, and detector stability under neural codec attacks.
\end{itemize}

\section{Related Work}

Audio watermarking embeds recoverable information while preserving perceptual quality. Compared with traditional spread-spectrum or echo-hiding methods~\cite{bender1996techniques,gruhl1996echo}, neural systems learn an embedder and detector jointly, optimizing imperceptibility, payload recovery, and robustness. Methods such as WavMark, AURA, SilentCipher, and related frameworks show strong robustness to common signal-processing attacks with extensive augmentation~\cite{chen2023wavmark,AURA,sanroman2024audioseal,singh2024silentcipher}. However, most still embed watermarks in waveform or time-frequency representations, making them vulnerable to neural codec pipelines that re-encode, quantize, and regenerate audio.

\subsection{Neural Audio Codecs}

Neural audio codecs are now widely used for compression and speech generation. EnCodec uses a SEANet-style encoder-decoder with an RVQ bottleneck to convert continuous acoustic features into compact discrete codes~\cite{defossez2022encodec}, while HiFi-Codec improves this paradigm with group-residual vector quantization~\cite{yang2023hificodec}. Such discrete representations are also used as intermediate tokens in speech synthesis and audio language modeling~\cite{wang2023valle,zhang2023speechtokenizer}. However, the same analysis-quantization-synthesis process can severely disrupt post-hoc watermarks.

\subsection{Codec-Robust Watermarking}
AudioSeal is one of the most representative recent neural audio watermarking systems for proactive voice-cloning detection~\cite{sanroman2024audioseal}. It jointly trains a watermark generator and a localized detector, and introduces a perceptual loss inspired by auditory masking to improve imperceptibility. Its codec-like design and augmentation strategy give it a certain degree of robustness against EnCodec-style processing, which motivates our use of a SEANet/EnCodec-like backbone. Although its released model also supports an optional 16-bit message, AudioSeal is primarily a watermark-presence and localization detector: its main metrics are detection accuracy, true-positive rate, false-positive rate, and AUC. We therefore treat AudioSeal as a strong but not directly equivalent related system rather than as a direct main-table baseline.

Our investigation differs from AudioSeal in the location and role of the codec-like structure. AudioSeal uses an encoder-decoder-style neural generator to synthesize a watermark mask or signal, which is then added to the waveform. The codec-like architecture is therefore used as a powerful watermark generator, while the actual embedding still occurs in the audio domain. In contrast, we use a codec-like autoencoder as the carrier pathway itself: the input waveform is first encoded into a continuous latent representation, the message-conditioned watermark is injected directly into this pre-decoder latent, and the decoder reconstructs speech from the modified latent. This changes the watermarking problem from waveform perturbation generation to latent representation modification. The goal of this paper is not to replace AudioSeal as a general-purpose watermarking system, but to investigate whether embedding in continuous pre-decoder latents can reduce the mismatch between watermark insertion and codec-style distortion, without assuming access to or modification of discrete codec tokens.

VoiceMark is closely related because it embeds watermarks into speaker-specific latents for zero-shot voice-cloning resistance~\cite{li2025voicemark}. It uses a pretrained RVQ tokenizer to separate content and speaker-specific components, embeds the watermark into the latter, and trains a transformer decoder with voice-cloning-simulated augmentations. This effectively bypasses RVQ disruption by selecting a stable latent carrier. However, VoiceMark targets voice-cloning transfer with a tokenizer-centered framework, whereas we study codec robustness for general deep-learning-based audio watermarking without assuming a specific pretrained tokenizer or fixed RVQ decomposition.

\section{Method}

\begin{figure*}[t]
    \centering
    \vspace{-0.9cm}
    \includegraphics[width=0.98\textwidth,height=0.45\linewidth,keepaspectratio=false]{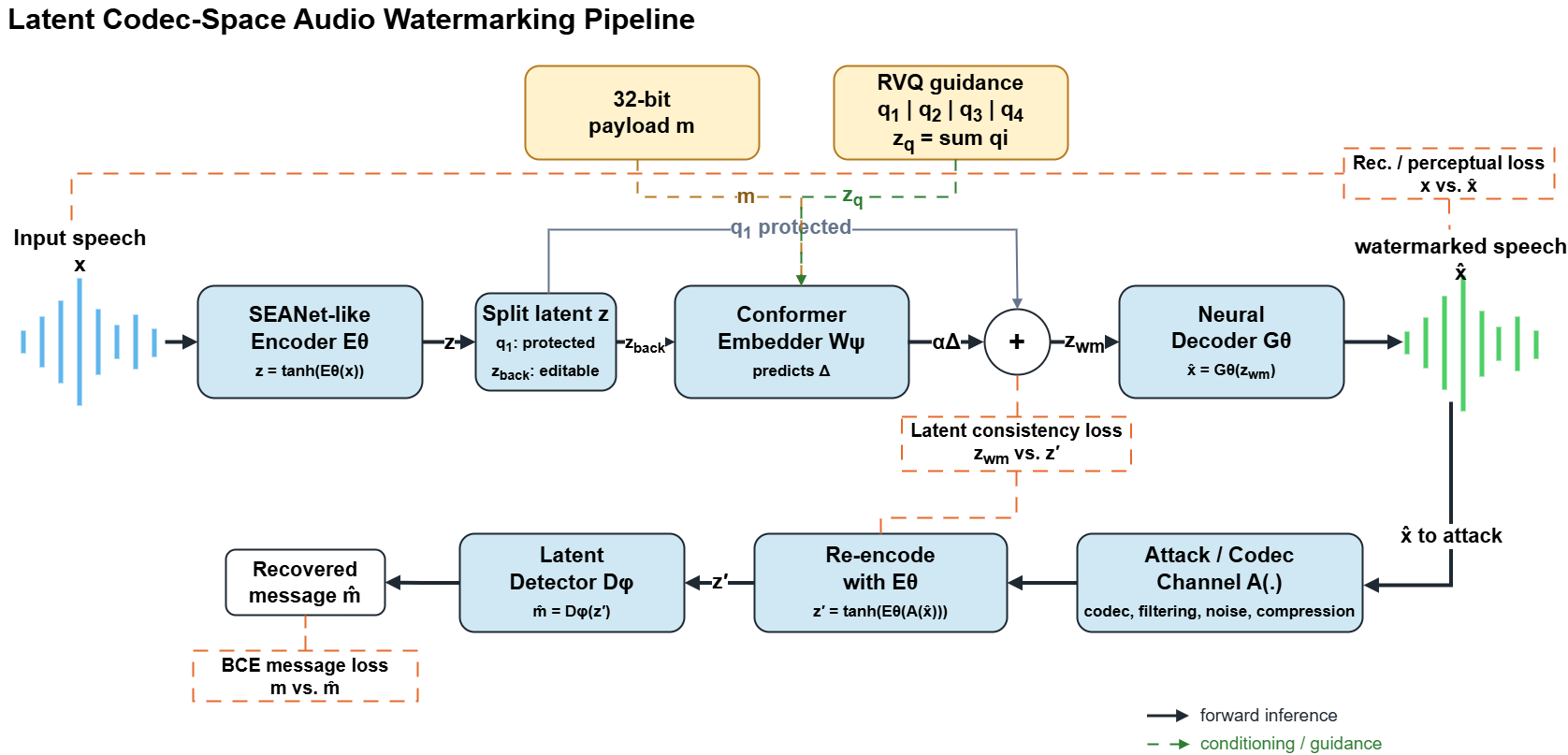}
    \caption{Overview of the proposed continuous latent-space audio watermarking
    pipeline.}
    
    \label{fig:latent_pipeline}
    \vspace{-0.6cm}
\end{figure*}

Figure~\ref{fig:latent_pipeline} gives an overview of the proposed continuous
latent-space watermarking pipeline.

\subsection{Notation and Definitions}

We use $x \in \mathbb{R}^{1 \times T}$ to denote the original speech waveform
and $m \in \{0,1\}^{K}$ to denote the binary payload. In all main experiments,
$K=32$. The encoder and decoder of the codec-like autoencoder are denoted by
$E_{\theta}$ and $G_{\theta}$, respectively. The continuous latent
representation is
\begin{equation}
    z = \tanh(E_{\theta}(x)).
\end{equation}
The hyperbolic tangent bounds the latent magnitude and keeps the watermark
perturbation in a stable range.

The RVQ module produces a sequence of residual components
$q_1,\ldots,q_N$, whose sum is denoted by
\begin{equation}
    z_q = \sum_{i=1}^{N} q_i .
\end{equation}
In this work, $z_q$ is a guidance representation for watermark generation and
front-layer decomposition; it is not the final latent forced into the decoder.
The first $M$ RVQ residual components form a protected front part
$z_{\mathrm{front}}=\sum_{i=1}^{M}q_i$, and the remaining continuous component
is written as $z_{\mathrm{back}}=z-z_{\mathrm{front}}$. A message-conditioned
embedder predicts a bounded residual $\Delta$, and the watermarked continuous
latent is denoted as $z_{\mathrm{wm}}$. The decoded watermarked waveform is
$\hat{x}=G_{\theta}(z_{\mathrm{wm}})$.

We use $\mathcal{A}(\cdot)$ for attack or codec transformations, including both
signal-processing operations and neural audio codecs. The detector $D_{\phi}$
predicts the payload from an attacked waveform after re-encoding. In the
experiments, \emph{watermarked} denotes the normal output $\hat{x}$,
\emph{no-embed} denotes reconstruction after removing the inference-time
watermark residual, and \emph{codec-only} denotes passing clean audio through a
codec without any watermark. The main reported checkpoints share the same
SEANet-based latent embedding architecture and differ mainly in the training
distribution, especially the emphasis placed on EnCodec-24k reconstruction.

\subsection{Problem Setup}

Given an input speech waveform $x \in \mathbb{R}^{1 \times T}$ and a binary watermark message $m \in \{0,1\}^{K}$, our goal is to generate a watermarked waveform $\hat{x}$ that remains perceptually close to $x$ while allowing the embedded message to be recovered after signal-processing and neural-codec transformations. We denote an attack or distortion operator as $\mathcal{A}(\cdot)$. The possibly attacked waveform is re-encoded before detection, and the recovered message is obtained as
\begin{equation}
    \hat{m} = D_{\phi}(E_{\theta}(\mathcal{A}(\hat{x}))).
\end{equation}
The training objective balances perceptual reconstruction quality against message recovery accuracy under $\mathcal{A}$.

Unlike waveform-domain watermarking, our method embeds the message in the continuous latent representation of a codec-like neural autoencoder. A SEANet-style encoder $E_{\theta}$ maps speech into a continuous latent representation
\begin{equation}
    z = \tanh(E_{\theta}(x)), \qquad z \in \mathbb{R}^{C \times L}.
\end{equation}
In the main system, $C=128$ and the total downsampling factor is $320$. A message-conditioned latent embedder $W_{\psi}$ predicts a bounded perturbation, and a decoder $G_{\theta}$ reconstructs the waveform from the watermarked latent.

\subsection{Latent Watermark Embedding}

The embedder is a message-conditioned deep latent transformation module. The binary message is first projected into an embedding, which conditions a stack of Conformer blocks through FiLM-style feature modulation~\cite{gulati2020conformer,FiLM}. The latent sequence is projected into a hidden dimension, processed by the message-conditioned blocks, and projected back to the original latent dimension. The final output is a bounded residual perturbation
\begin{equation}
    \Delta = \tanh(W_{\psi}(u,m)),
\end{equation}
where $u$ is the latent feature used to condition the embedder.

Earlier versions also explored multiplicative latent masking,
\begin{equation}
    z_{\mathrm{wm}} = z \odot M_{\psi}(z,m),
\end{equation}
where the message-conditioned mask directly rescales the latent channels. This
form was unstable in our experiments because the watermark operation was not
localized to a small residual direction: even a mask close to one changes the
relative magnitude of many content-bearing latent coefficients. After neural
decoding, these scale changes could produce audible waveform distortion and
large reconstruction loss. The main reported variants therefore use additive
residual injection:
\begin{equation}
    z_{\mathrm{wm}} = z_{\mathrm{base}} + \alpha \Delta,
\end{equation}
where $\alpha$ is a small learned scale. This form makes the watermark strength easier to control and is the basis of the reported variants.

\subsection{RVQ-Guided Latent Decomposition}

Residual vector quantization~\cite{zeghidour2021soundstream,defossez2022encodec} is used as a guidance and decomposition mechanism, not as the mandatory reconstruction bottleneck of the final model. Given $z$, a stack of RVQ layers produces residual components $q_1,\ldots,q_N$ and their sum
\begin{equation}
    z_q = \sum_{i=1}^{N} q_i .
\end{equation}
In an early design, RVQ was placed directly between the encoder and decoder so
that reconstruction had to pass through quantized watermarked latents. This can
be viewed as an attempt to move from continuous latent-space embedding toward a
more discrete codec-space embedding, where the watermark must survive codebook
assignment before waveform synthesis. This hard quantized path was difficult to
optimize for two reasons. First, nearest codebook assignment could absorb or
erase small watermark perturbations before they reached the decoder. Second, the
decoder had to reconstruct speech from discontinuous codebook entries while the
detector still required message-specific information to survive. In preliminary
runs, this design failed to converge to reliable payload recovery, with bit
accuracy remaining below about $70\%$, so we removed RVQ from the mandatory
reconstruction path. This motivated the reported continuous latent-space design:
the watermark is inserted before neural decoding, while RVQ is retained only as
guidance for how the continuous latent relates to codec-like quantization.

In our main investigated configuration, reconstruction is kept continuous. RVQ serves two roles. First, $z_q$ can be used as the conditioning input to the watermark embedder, so that the predicted perturbation is guided by a codec-like decomposition of the latent. Second, the residual components define front and back latent parts:
\begin{equation}
    z_{\mathrm{front}} = \sum_{i=1}^{M} q_i, \qquad
    z_{\mathrm{back}} = z - z_{\mathrm{front}}.
\end{equation}
The front components are treated as more perceptually important and can be protected from direct watermark modification. This is a continuous decomposition of $z$, not a hard quantized decoder input. With front-layer protection, the watermarked latent is
\begin{equation}
    z_{\mathrm{wm}} =
    z_{\mathrm{front}} +
    \left(z_{\mathrm{back}} + \alpha \Delta_{\mathrm{back}}\right).
\end{equation}
When projection is enabled, the predicted perturbation is projected away from the protected front component before being added. Thus, RVQ does not replace the continuous latent passed to the decoder; it guides where and how the watermark is injected.

\subsection{Codec-like Reconstruction and Detection}
\label{codeclike}
The decoder synthesizes the watermarked waveform as
\begin{equation}
    \hat{x} = G_{\theta}(z_{\mathrm{wm}}).
\end{equation}
Because the waveform is generated from a modified continuous latent representation, the perceptual quality is influenced both by the watermark perturbation and by the reconstruction capacity of the encoder-decoder itself.

For detection, the possibly attacked waveform is re-encoded:
\begin{equation}
    z' = \tanh(E_{\theta}(\mathcal{A}(\hat{x}))).
\end{equation}
A latent detector predicts $K$ message logits from $z'$. We report bit accuracy. Bit accuracy denotes \textbf{aggregate bit recovery rate}, i.e., 1-BER, averaged over all payload bits and utterances.

\subsection{Training Pipeline}

Training is staged. The early stage emphasizes reconstruction and proxy latent detection, where the detector observes the watermarked latent directly with latent perturbations. This stabilizes message learning before the full waveform path is reliable. Later stages introduce real waveform reconstruction, re-encoding, and attack simulation, forcing the detector to recover the message from $E_{\theta}(\mathcal{A}(\hat{x}))$.

For the EnCodec-24k-focused and EnCodec-heavy regimes, training increases
exposure to EnCodec-24k in different ways. The focused regime does not choose
EnCodec more often; after EnCodec is chosen, it removes the 16 kHz stress route.
The heavy endpoint additionally raises the EnCodec sampling probability. These
regimes use a straight-through estimator~\cite{bengio2013estimating} for codec
simulation: the forward pass exposes the detector to codec-reconstructed audio,
while the backward pass approximates the codec transformation with an identity
gradient. The balanced training regime keeps the same latent watermarking
structure but does not make EnCodec-24k the dominant training objective.

As shown in Fig.~\ref{fig:latent_pipeline}, the pipeline encodes speech into
$z$, uses RVQ for guidance and front-layer protection, predicts a
message-conditioned residual, decodes the watermarked latent, and re-encodes
the attacked waveform for latent-domain detection. The shared encoder couples
reconstruction quality with detection robustness.

\subsection{Loss Function}

The generator is trained with reconstruction, perceptual, latent-consistency, adversarial, RVQ, and message losses:
\begin{align}
\mathcal{L}_{G} =
& \lambda_{\mathrm{mel}}\mathcal{L}_{\mathrm{mel}}
+ \lambda_{\mathrm{stft}}\mathcal{L}_{\mathrm{MRSTFT}}
+ \lambda_{\mathrm{tf}}\mathcal{L}_{\mathrm{TF}}
+ \lambda_{\mathrm{adv}}\mathcal{L}_{\mathrm{adv}} \nonumber \\
&+ \lambda_{\mathrm{sim}}\|z_{\mathrm{wm}} - z\|_2^2
+ \lambda_{\mathrm{lat}}\|E_{\theta}(\mathcal{A}(\hat{x})) - z_{\mathrm{wm}}\|_2^2 \nonumber \\
&+ \lambda_{\mathrm{msg}}\mathcal{L}_{\mathrm{BCE}}(D_{\phi}(E_{\theta}(\mathcal{A}(\hat{x}))),m)
+ \lambda_{\mathrm{vq}}\mathcal{L}_{\mathrm{RVQ}} .
\end{align}
The reconstruction term $\mathcal{L}_{\mathrm{mel}}$ is a multi-scale
mel-spectrogram loss. It compares the original and reconstructed waveforms at
multiple FFT scales using both linear mel distance and log-mel distance. This
term provides a broad spectral constraint and is the main reconstruction loss
used throughout training.

The multi-resolution STFT loss $\mathcal{L}_{\mathrm{MRSTFT}}$ is the average
$\ell_1$ distance between log-magnitude STFTs of $x$ and $\hat{x}$ at three
$(n_{\mathrm{fft}},h,w)$ settings: $(1024,120,600)$, $(2048,240,1200)$, and
$(512,50,240)$. This loss penalizes spectral artifacts not fully captured by mel
features.

The time-frequency loudness ratio loss $\mathcal{L}_{\mathrm{TF}}$ splits the
waveform into frequency bands and short temporal frames, then compares the
loudness of the reconstruction error with the loudness of the reference signal.
It gives larger weight to distortions that are more likely to be perceptually
salient. The adversarial term $\mathcal{L}_{\mathrm{adv}}$ is computed with
multi-period and multi-scale discriminators~\cite{kong2020hifigan}, including both generator
adversarial loss and feature matching.

The message loss is binary cross-entropy. During early training, a proxy branch
applies BCE to the detector output from the watermarked latent. During later
training, the real branch applies BCE after waveform reconstruction, attack
simulation, and re-encoding:
\begin{equation}
    \mathcal{L}_{\mathrm{msg}}
    = \mathrm{BCE}(D_{\phi}(E_{\theta}(\mathcal{A}(\hat{x}))),m).
\end{equation}
The latent-consistency loss encourages the attacked and re-encoded latent to
remain close to the watermarked latent, while $\mathcal{L}_{\mathrm{sim}}$
limits deviation from the original latent for perceptual preservation.
$\mathcal{L}_{\mathrm{RVQ}}$ contains the standard codebook and commitment
losses for stabilizing the guidance quantizers. Different variants share this
objective family but use different weights.

\subsection{Backbone Controls}

The main method uses the SEANet-style encoder-decoder. To analyze whether perceptual quality is limited by watermark perturbation or by the reconstruction backbone itself, we also run diagnostic backbone controls. One control trains the same SEANet-style autoencoder without watermarking and without RVQ/codebook loss. A second control replaces the SEANet-style encoder and decoder with a trainable DAC-style residual convolutional autoencoder~\cite{kumar2023dac} while keeping the same latent dimensionality and downsampling factor. This DAC-style model is not the pretrained DAC codec; it is a local backbone-control architecture. These DAC-related experiments are diagnostic controls and are not treated as the main watermarking method.

\section{Experiments}
\label{sec:experiments}

\subsection{Experimental Setup}
\label{subsec:setup}

All experiments use speech segments from the English subset of Emilia, an extensive multilingual and diverse speech dataset for large-scale speech generation~\cite{he2024emilia}. Audio is resampled or loaded at $48$ kHz and cropped or padded to a maximum length of $120000$ samples. The payload length is $K=32$ bits. Unless otherwise stated, the main SEANet backbone uses $C=128$ latent channels and downsampling ratios $[8,5,4,2]$, resulting in a total downsampling factor of $320$. The main RVQ-guided variants use four RVQ groups with a codebook size of $1024$ and protect the first RVQ residual component from direct watermark modification.

\subsection{Evaluation Protocol and Metrics}
\label{subsec:metrics}

For watermark robustness evaluation, a payload is embedded into each speech
segment, the watermarked waveform is optionally transformed by an attack
$\mathcal{A}$, and the detector predicts the payload from the attacked waveform.
We report bit accuracy as detailed in Section \ref{codeclike}. We also report BCE as an
optimization diagnostic.

For perceptual quality, we report PESQ~\cite{itu2001pesq} as the main objective speech-quality
metric and ViSQOL~\cite{Chinen20_ViSQOL} as a complementary full-reference perceptual score on the same 1000-sample subset.
We also report STOI~\cite{taal2011stoi} to reflect intelligibility and SNR/SI-SDR~\cite{leroux2019sdr} as
waveform-level distortion references. SNR and SI-SDR are useful for tracking
signal distortion, but they do not always align with perceived quality;
therefore, quality discussions primarily rely on PESQ together with listening
checks when available \cite{AURA}. For PESQ/STOI evaluation, signals are resampled to the metric-required sampling rate before scoring like MaskMark \cite{maskmark}.

The attack suite contains common signal-processing attacks and neural-codec
attacks. The signal-processing group includes additive noise, sample
suppression, low-pass filtering, high-pass filtering, band-pass filtering,
resampling, amplitude scaling, quantization, smoothing, pink noise,
reverberation, time stretching, random phase shift, spectrogram augmentation,
and segment replacement. The neural-codec group mainly focuses on EnCodec; HiFiCodec is used as an additional codec-quality reference.
For EnCodec, we report two reconstruction conditions. The main codec condition
uses the official 24 kHz EnCodec model and is denoted EnCodec-24k. Two 24 kHz
configurations were evaluated during development, but they produced the same
reconstructed waveform in our evaluation pipeline; we therefore merge them into
one EnCodec-24k result rather than presenting a bitrate sweep. We additionally
include a stronger stress condition, EnCodec-16k, which first resamples the
audio to 16 kHz before EnCodec reconstruction.

\subsection{Training Regimes}
\label{subsec:variants}

The principal SEANet-based checkpoints use the same latent embedding
architecture: $z_q$ guides the watermark embedder, the first RVQ residual
component is protected from direct modification, and the decoder receives a
continuous watermarked latent. Their difference is therefore procedural rather
than architectural.

The balanced training regime uses the default attack sampler and does not make
EnCodec-24k the dominant condition. When an EnCodec attack is selected, the
sampler includes both the 16 kHz stress route and the 24 kHz EnCodec routes.
The EnCodec-24k-focused regime keeps the same finetuning loss coefficients but
restricts the EnCodec branch to the 24 kHz routes, so that codec-reconstruction
updates are focused on EnCodec-24k. The EnCodec-heavy endpoint keeps the same
finetuning loss coefficients and 24 kHz EnCodec routes, but further biases the
attack sampler toward EnCodec relative to clean/origin examples. Thus, these
regimes differ mainly in the attack sampling distribution rather than in the
scalar loss weights. 

In implementation, the attack sampler assigns each candidate attack an
unnormalized sampling weight and then normalizes these weights into
probabilities. In the EnCodec-24k-focused branch, the EnCodec branch is restricted to
the two 24 kHz EnCodec routes and samples them with equal weights. With the
default attack list, this corresponds to an EnCodec attack probability of
$1/17=5.9\%$, split equally across the two 24 kHz routes. In the EnCodec-heavy
endpoint, EnCodec receives a larger sampling weight of $2.5$, while the
clean/origin case receives weight $0.8$ and the remaining attacks keep weight
$1.0$. The total sampling weight is therefore $18.3$, so EnCodec is selected
with probability $2.5/18.3=13.7\%$; each 24 kHz route has probability $6.8\%$.
This changes how often codec reconstruction is sampled without changing the
scalar loss coefficients.

The DAC-style full-pipeline probe is reported separately as a backbone-control
experiment: it keeps the watermarking pipeline fixed but replaces the SEANet
autoencoder with a trainable DAC-style autoencoder.

\subsection{Reconstruction Diagnostics}
\label{subsec:ceiling}

We separate the effect of the inference-time watermark residual from the
reconstruction behavior of the jointly trained autoencoder. For a trained
watermarking model, the \emph{no-embed} condition removes the watermark delta at
inference time:
\begin{equation}
    x_{\mathrm{noembed}} = G_{\theta}(\tanh(E_{\theta}(x))).
\end{equation}
This comparison is not a pure autoencoder ceiling, because the encoder and
decoder have already been optimized together with watermark and robustness
objectives. It is still useful as a local diagnostic: in a 1000-sample
diagnostic, removing the residual barely changes PESQ under balanced training
(3.734 without the residual versus 3.727 with the residual), while under
EnCodec-24k-focused training the no-embed output is lower than the watermarked output
(3.426 versus 3.514).
Thus, the inference-time residual alone does not explain perceptual quality.
The autoencoder weights, latent residual, and detector/attack objectives are
coupled. To estimate the reconstruction behavior more cleanly, we therefore
train watermark-free autoencoders in the backbone probes below.

\subsection{Codec Reconstruction Reference}
\label{subsec:codec_ceiling}

We also pass clean audio through neural codecs without watermarking to
contextualize the quality of attacked audio. On 1000 samples, clean audio obtains
PESQ $4.644$. EnCodec-24k reconstruction obtains PESQ $3.626$, while the
HiFiCodec-24k variants tested in our current pipeline obtain PESQ around
$2.5$--$2.6$. These numbers are not claims about the intrinsic upper bound of
each codec under all configurations; they provide a reference for the specific
codec conditions used in our attack evaluation. As described above, EnCodec-24k
is treated as a fixed reconstruction condition rather than as a bitrate sweep.

\subsection{Robustness Results}
\label{subsec:robustness}

Table~\ref{tab:main_tradeoff} summarizes the quality-robustness trade-off caused by increasing the EnCodec-24k emphasis during training. All accuracy columns report bit accuracy. Balanced training achieves the highest clean perceptual quality but lower EnCodec-24k bit accuracy. EnCodec-24k-focused training substantially improves EnCodec-24k robustness with a moderate PESQ reduction and is therefore used as the main codec-aware setting. The EnCodec-heavy endpoint provides a further robustness gain but costs additional PESQ. ViSQOL follows the same ordering but with smaller separation, so we keep PESQ as the primary quality metric.

\begin{table}[t]
\vspace{-0.4cm}
\centering
\caption{Quality and codec robustness trade-off on 1000 samples.}
\label{tab:main_tradeoff}
\resizebox{\columnwidth}{!}{%
\begin{tabular}{lccccc}
\hline
Variant & Step & PESQ & ViSQOL & Clean Acc. & EnCodec-24k Acc. \\
\hline
Balanced training & 258k & 3.727 & 4.590 & 99.8\% & 78.8\% \\
EnCodec-24k-focused training & 258k & 3.514 & 4.553 & 100.0\% & 95.6\% \\
EnCodec-heavy endpoint & 222k & 3.427 & 4.531 & 100.0\% & 97.1\% \\
\hline
\end{tabular}
}
\end{table}

Under non-codec signal-processing attacks, the EnCodec-24k-focused checkpoint also remains strong. For example, at 258k it achieves bit accuracy above $98\%$ for noise, suppression, low-pass filtering, high-pass filtering, band filtering, resampling, amplitude scaling, quantization, smoothing, pink noise, stretching, spectrogram augmentation, and replacement. However, EnCodec-16k remains highly destructive for all current checkpoints, with bit accuracy close to chance. This indicates that robustness to EnCodec-24k does not automatically transfer to the more severe EnCodec-16k condition.

\subsection{Backbone Diagnostics}
\label{subsec:backbone_probe}

To test whether perceptual quality is limited only by the watermark perturbation
or also by the reconstruction backbone, we run several diagnostic backbone
probes:
\begin{itemize}
\item SEANet plain AE: current SEANet autoencoder, no watermark, no RVQ/codebook loss.
\item DAC-style plain AE: trainable DAC-style autoencoder, no watermark, no RVQ/codebook loss.
\item DAC-style full pipeline: proposed watermarking pipeline with DAC-style encoder/decoder replacing SEANet.
\end{itemize}

The two plain AE probes test the reconstruction behavior of the backbone without
watermark objectives. The DAC-style full-pipeline probe tests whether replacing
the backbone inside the watermarking pipeline preserves both audio quality and
payload recovery.

Table~\ref{tab:backbone_probe} reports the latest available probe results. All
accuracy columns report bit accuracy. The
DAC-style plain autoencoder obtains slightly higher PESQ and ViSQOL than the SEANet plain
autoencoder, which motivates testing whether a stronger standalone
reconstruction backbone also improves the full watermarking pipeline. However,
the DAC-style full-pipeline checkpoint does not improve PESQ or codec robustness
over the main SEANet-based checkpoints. This result should be interpreted as a
compatibility diagnostic, not as a final judgment on DAC-style backbones. It
suggests that latent watermarking depends on the joint behavior of the embedder,
detector, attack training, and reconstruction backbone, rather than on plain AE
reconstruction capacity alone.

\begin{table}[t]
\vspace{-0.4cm}
\centering
\caption{Backbone probe results on 1000 samples.}
\label{tab:backbone_probe}
\resizebox{\columnwidth}{!}{%
\begin{tabular}{lccccc}
\toprule
Probe & Step & PESQ & ViSQOL & Clean Acc. & EnCodec-24k Acc. \\
\midrule
SEANet plain AE & 120k & 3.792 & 4.606 & -- & -- \\
DAC-style plain AE & 120k & 3.838 & 4.636 & -- & -- \\
DAC-style full pipeline & 276k & 3.419 & 4.566 & 95.8\% & 72.5\% \\
\bottomrule
\end{tabular}
}
\end{table}

\section{Limitations}

This work is an investigation of continuous latent-space watermark embedding
rather than a new universal watermarking baseline. Although EnCodec-24k
robustness improves, EnCodec-16k remains highly destructive, with bit accuracy
close to chance. We also do not embed messages into discrete codec tokens or RVQ
indices; instead, we study how far a continuous pre-decoder latent carrier can
go under neural-codec reconstruction.

\section{Conclusion}

This paper investigates codec-robust audio watermarking by moving message
embedding into continuous latent representations of codec-like autoencoders. The results show a clear
trade-off: codec-aware training improves EnCodec-24k robustness, while
perceptual quality remains coupled to the watermark objective and autoencoder
reconstruction behavior. Backbone probes further show that stronger standalone
autoencoding is not sufficient; the backbone must also remain compatible with
message embedding, detection, and attack-aware training.

\section{Generative AI Use Disclosure}
Large Language Models (LLMs) were used only for manuscript polishing, such as
rephrasing and grammar checks. They were not used for ideation, methodology,
experimental design, data analysis, or result interpretation. All scientific
content was produced and verified by the authors.

\bibliographystyle{IEEEtran}
\bibliography{refs}

\end{document}